# Simultaneous Addition of $B_4C$ + SiC to $MgB_2$ Wires and Consequences for $J_c$ and $B_{irr}$

René Flükiger[1,2,*], Paola Lezza[1], Marco Cesaretti[3], Carmine Senatore[1] and Roman Gladyshevski[4]

[1] Group of Applied Physics, University of Geneva, 20, rue de l'Ecole de Médecine – CH-1211 Geneva 4, Switzerland

[2] Department of Physics of Condensed Matter (DPMC) and MANEP (NCCR), University of Geneva, 24, quai Ernest Ansermet – CH-1211 Geneva 4, Switzerland

[3] Dipartimento di Matematica Fisica, Univ. Cattolica del Sacro Cuore, Brescia

[4] Department of Inorganic Chemistry, University of Lviv, Ukraine

**Abstract**

The simultaneous addition of various Carbon based additives (in the present case $B_4C$ + SiC) to Mg and B powders has been introduced as a new concept in view of enhancing the superconducting parameters $B_{c2}$, $B_{irr}$ and $J_c$ values of *in situ* Fe/$MgB_2$ wires. A series of Fe sheathed monofilamentary wires of 1.1 mm diameter with a $MgB_2$ core of 600 μm diameter was prepared with various $B_4C$:SiC ratios, the relation being 2.5:7.5, 5:5, 7.5:2.5 and 7.5:7.5 (values in wt. %). After reaction of 1 hour at 760°C, the wire containing 7.5 wt.% $B_4C$ and 2.5 wt.% SiC powders exhibited a $J_c$ value of $1 \times 10^4$ A/cm$^2$ at 11.3 T and 4.2 K. Although only 2.5 wt.% SiC were added, these values are considerably above those of ternary wires with $B_4C$ additions, where the same $J_c$ value is obtained at 10T. The slope $J_c$ vs. B for the $B_4C$ + SiC wires is steeper than for SiC additives, the $J_c$ values at 4.2 K being superior at fields below 9 T. The lattice parameters *a* of the $B_4C$ + SiC added wires exhibit lower values than ternary wires with the same nominal C content, suggesting a higher C content in the $MgB_2$ phase. The "disorder" in the $MgB_2$ structure has been characterized as a partial substitution of B by Carbon. A reduction of the domain sizes in the *c* direction as well as in-plane to 250 Å and 124 Å has been determined from the FWHM and the breath values of the (110) and (002) peaks. With the simultaneous introduction of $B_4C$ + SiC, a strong improvement of $J_c$ and $B_{irr}$ has been obtained with respect to $B_4C$ additions.

A further enhancement of $J_c$ is expected when using different combinations of additives with and without Carbon, aiming for a further raise of $J_c$ in wires with multiple additives, as a result of the combination of different mechanisms.

* Corresponding author; *e-mail address*: rene.flukiger@physics.unige.ch; *tel.*: +41 22 3796240

**Introduction**

With the progress of the superconducting current carrying capability of $MgB_2$ wires since its discovery in 2001 [1], the question arises whether this compound can in some particular cases be considered as a possible substitute for NbTi or $Nb_3Sn$. The positive arguments for $MgB_2$ are its high transition temperature and its weak-link free character, the low material costs and the small anisotropy. $MgB_2$ appears as a promising candidate for engineering applications, e.g. MRI magnets at 20 K and intermediate field inserts for NMR magnets. However, the superconducting parameters $B_{c2}$, $B_{irr}$ and $J_c$ need to be further increased. The present paper is centered on the *in situ* technique, which allows the addition of a great variety of elements and compounds, in view of the optimization of the critical current density by the combination of different mechanisms.

*The additives to $MgB_2$*

An impressive amount of works has been published about the effect of additives on the critical parameters of $MgB_2$ wires and tapes. The most efficient additive found so far is SiC [2-6], for which a $J_c$ (4.2K) value of $1 \times 10^4$ A/cm$^2$ has been reported [2] for fields slightly above 12 T at 4.2 K. Other known additives are C (as nanotubes or diamond powder) [7,8], WB [9], $WSi_2$, $ZrSi_2$ [3,10], Ti [11], $Si_3N_4$ [12,13], $Na_2CO_3$ [14], aromatic hydrocarbons [15], hydrides [5] and $B_4C$ [16,17,20]. Most of these very different additives lead to an enhancement of $J_c$, suggesting that more than one mechanism must affect the critical parameters of $MgB_2$ wires.

As possible mechanism one can imagine the following ones: the partial substitution of Boron by Carbon or of Mg by other elements, a higher densification of the $MgB_2$ filaments during reaction, the reduction of the $MgB_2$ domains (i.e. of cristallinity) and the effect of secondary phases on the conditions at the grain boundaries. The correct attribution of the mechanism responsible for the $J_c$ enhancement is not easy, due to the superposition of various mechanisms. So far, a main source for the enhancement of $B_{c2}$ and $B_{irr}$ has been identified as the enhancement of the normal resistivity, due to the partial substitution of Boron by Carbon in the $MgB_2$ structure.

In contrast to bulk samples, no gases should be released during the reaction in long, sheathed $MgB_2$ wires. This excludes hydrides, hydrocarbons or nitrides as potential additives. In addition, the *in situ* reaction temperature of the powder mixture with the additive should be as low as possible, to prevent a reaction with the sheath materials, but also to prevent a too important grain growth. It is known that the reaction temperature for optimized $J_c$ values after Carbon additives is well above 900°C [18,19]. Yamamoto et al. [16] have shown that the reaction with $B_4C$ occurs at lower temperatures: they reacted bulk samples at 850°C and obtained at 4.2 K the value of $J_c = 1 \times 10^4$ A/cm$^2$ above 8 T. It was recently shown in our laboratory (Lezza et al. [17]) that this field could be raised to 9 T after reaction temperatures as low as 800°C. In addition to the partial substitution of B by C, a correlation between the cristallinity and $B_{irr}$ was proposed by Yamamoto et al. [20] and Kumakura et al [21] for both, SiC and $B_4C$ additives to $MgB_2$, on the basis of the FWHM values of the (110) peaks in X-ray patterns.

The present work is a first attempt to introduce simultaneously various Carbon containing additives, following the goal of an enhanced presence of Carbon on the Boron sites, hoping for a further optimization of the superconducting parameters. Since the slope of $J_c$ vs. $B$ for $B_4C$ added *in situ* wires is higher than for SiC additions [17], it was hoped that slightly different mechanisms could have a positive influence on the superconducting properties of the quaternary wire.

**Experimental**

Monofilamentary wires were prepared by the *in situ* technique, using Mg (99.8% pure, ~325 mesh) and amorphous B (1.02 % oxygen, submicrometric particle size) at the stoichiometric ratio, adding various amounts of $B_4C$ powders (particle size: 500 nm) or both $B_4C$ and SiC at the ratios listed in Table 1. The precursor powders were mixed and then inserted in Fe tubes of $\varnothing_{out}$ and $\varnothing_{in}$ of 8 and 5 mm, respectively. The diameter of the tubes was reduced applying a combination of swaging and drawing to final diameters of 1.11 and 0.657 mm, respectively. The wires were then heat treated under Ar flow at various temperatures up to 800°C for 1 hour. For comparison, undoped wires were fabricated with the same procedure up to diameters of 1.11 mm diameter and annealed both at 670 °C, 2 min or 800°C, 1 hour [17].

The critical currents were measured by the standard 4-point method at fields up to 15 T, both at 4.2 K and 20 K. X ray diffraction measurements were performed after extracting the powder from the filaments, using Cu $K\alpha_1$ radiation with a Si standard for a precise determination of the lattice parameter changes. $T_c$ was measured with a SQUID magnetometer in a background field of 12.3 Oe, ramping from 20 to 40 K.

**Results**

The wires studied in the present work are characterized in Table I. Wire A consists of binary $MgB_2$, while B, C and D are ternary wires with 5, 10 and 12.5 wt.% $B_4C$. The quaternary wires E,F,G and H contain simultaneously $B_4C$ and SiC additives, their respective amounts being 7.5 and 2.5 wt.%, 5 and 5 wt.%, 2.5 and 7.5 wt.%, and 7.5 and 7.5 wt.% with respect to $MgB_2$. The lattice parameters of the samples studied in the present work are listed in Table II. The parameters *a* and *c* are plotted in Fig. 1 as a function of the nominal Carbon: Boron ratio, confirming that $c$ is little affected by the Carbon content, in contrast to $a$, which shows a gradual decrease.

The decrease of the lattice parameter $a$ had already been reported by several authors for SiC additions, while Yamamoto et al. [16,20], and more recently Lezza et al. [17] reported a decrease for $B_4C$ additions to $MgB_2$. The stars in Fig. 1 represent the values for the $B_4C$ additives, while the circles represent the data for simultaneous $B_4C$ + SiC addition.

Fig. 2 shows the transport critical current densities for the binary wire A and for the $B_4C$ added wires, B, C and D at 4.2 K vs. the applied field, after reaction at 800°C for 1 hour. The present wires had a $MgB_2$ core of 600 μm diameter, and contained neither a Nb barrier nor a Cu stabilizer, which considerably limited the measuring field range. The present data are thus incomplete, but allow a good comparison among the different additions, at least at fields around 10 T and 4.2 K.

After reaction at 800°C, the binary $MgB_2$ wire shows strongly enhanced values with respect to the optimized temperature, 670°C. For the $B_4C$ added wires, a maximum of $J_c$ was found at 10 wt.% $B_4C$: the $J_c$ value of $1 \times 10^4$ A/cm$^2$ is reached at 96 T [17]. After reaction at 760°C for 1 hour (not shown in Fig. 2), this value was raised to 10.0 T, the corresponding value at 20 K being 4.5 T.

The variation of $J_c$ vs. B for the quaternary, $B_4C$ + SiC added wires E, F, G and H is shown in Fig. 3 after 1 hour at 800 °C. The maximum values are reached for wire E with 7.5 wt.% $B_4C$ and 2.5 wt.% SiC, for which the value of $1 \times 10^4$ A/cm2 at 4.2 K is reached at 11.1 T. The data on wire E (7.5 wt.% $B_4C$ + 2.5 wt.% SiC) at 4.2 and 20K are plotted in Fig. 4, together with the data of Soltanian et al. [4]. A further improvement was obtained after 1 hour at 760°C, where $1 \times 10^4$ A/cm$^2$ at 4.2 K was reached at 11.3 T. It was shown by Lezza et al [17] that the slope at 4.2 K of $J_c$ vs. B for $B_4C$ added wires is considerably steeper than for SiC added wires. This tendency is also present in the $B_4C$ + SiC added wire E after 1 hour at 800°C and is even slightly enhanced after reaction at 760°C. It follows that $J_c$ of the SiC added wire is still higher at higher fields, but that there is a cross over in the region of 9 T.

From the resistive transitions, values of $B_{c2}$ and $B_{irr}$ have been extracted up to a field of 10 T (Fig 5). For all measured wires, an increase in the magnetic field caused a shift toward lower temperatures (Fig. 5). $B_{c2}$ and $B_{irr}$ values were defined using the criteria of 90% and 10%, respectively, of the normal state resistive curves. As shown in Fig.5, the values of $B_{c2}$ and $B{irr}$ at 20 K are 8 and 10 T, respectively. From a comparison with literature data, the extrapolation to lower temperatures yields roughly 27 and 23 T, respectively.

The $T_c$ values of the analyzed wires vary monotonously with the lattice parameter *a* (Fig. 6). Yamamoto et al. [20] and Kumakura et al. [21] reported that the addition of $B_4C$ and SiC, respectively, to $MgB_2$ leads to an increase of the FWHM values of the (110) and (200) peaks, thus reflecting a lower cristallinity. A correlation between the FWHM value and $B_{irr}$ was found, in particular for the (110) reflections.

The present data were analyzed by means of the DIFPATAN software, which takes into account the FWHM values as well as the integral breath $\beta = A/I(0)$, where A is the peak area and I(0) the peak height. Wire E with the highest $J_c$ values exhibited FWHM values for the (002) and (110) reflections of 0.41 and 0.62°, respectively, the corresponding values for the breath $\beta$ being 0.56 and 0.94. The calculation yielded domain sizes of 248 Å for the (002) peak and 124 Å for the (110) peak, which is considerably smaller than the corresponding values for the $MgB_2$ wire without additives.

The tendency of increasing FWHM values with increasing amount of additives confirms the results of Refs. [4,16] and [20]. The situation in doped $MgB_2$ wires can be described as follows. After reaction at 800°C: the grains, of a size of 300 – 500 nm [20] consist of domains having an average extension of 248 Å in the *c* direction, the in-plane size being 124 Å. There is an appreciable uncertainty in these values, due to the inhomogeneities of the Carbon substitution, which are difficult to evaluate. A systematical study will lead to a better characterization.

**Conclusions**

The simultaneous addition of various Carbon based additives (in the present case $B_4C$ + SiC) to Mg and B powders has been introduced as a new concept in view of enhancing the superconducting parameters $B_{c2}$, $B_{irr}$ and $J_c$ values of *in situ* Fe/MgB$_2$ wires. A series of Fe sheathed monofilamentary wires of 1.1 mm diameter with a MgB$_2$ core of 600 µm diameter was prepared with various $B_4C$:SiC ratios, the relation being 2.5:7.5, 5:5, 7.5:2.5 and 7.5:7.5 (values in wt. %). The highest value of $J_c$ was obtained for the wire containing 7.5 wt.% $B_4C$ and 2.5 wt.% SiC powders, reacted 1 hour at 760°C. A $J_c$ value of $1 \times 10^4$ A/cm$^2$ was measured at 11.3 T and 4.2 K.

Although only 2.5 wt.% SiC were added, these values are considerably above those of ternary wires with $B_4C$ additions, where this $J_c$ value is obtained at 10T. The lattice parameters *a* of the wires with quaternary additions ($B_4C$ + SiC) exhibit lower values than $B_4C$ added wires with the same nominal C content, suggesting a higher C content in the MgB$_2$ phase. With the simultaneous introduction of $B_4C$ + SiC, a strong improvement of $J_c$ and $B_{irr}$ has been obtained over binary *in situ* wires. The slope of $J_c$ vs. $B$ for the $B_4C$ + SiC added is steeper than for wires with SiC additives, resulting in higher $J_c$(4.2K) values for magnetic fields below 9 T.

The "disorder" in the MgB$_2$ structure has been characterized as a partial substitution of B by Carbon and by the reduction of the domain sizes in the *c* direction as well as in-plane, to values well below 250 Å, as determined from the FWHM and the breath values of the (110) and (002) peaks.

An enhancement of the present data is expected for Cu stabilized wires with higher reduction ratios and a Nb barrier [22] to minimize the reaction with the Fe sheath. A further enhancement of $J_c$ is expected when using different combinations of additives with and without Carbon, aiming for a further raise of $J_c$ in wires with multiple additives, as a result of the combination of different mechanisms.

**Acknowledgements**

The authors would like to thank Drs. W. Goldacker and Sonja Schlachter of the ITP at the Forschungszentrum Karlsruhe for many discussions about the *in situ* technique.

This work is supported by the EU FP 6 project NMP-3-CT2004-505724 and by MANEP (NCCR), Switzerland.

**Table Captions**

**Table 1:** Relative percentage of additive and nominal Carbon content with respect to the total Boron content.

**Table 2:** Nominal Carbon/Boron ratios and lattice parameters of the wires annealed at 800°C for 1 hour.

**Table 1**

| Sample | wt.% B$_4$C | wt.% SiC | Nominal C / B ratio |
|:---:|:---:|:---:|:---:|
| A | 0 | 0 | 0 |
| B | 5 | 0 | 0.019 |
| C | 10 | 0 | 0.036 |
| D | 12.5 | 0 | 0.043 |
| E | 7.5 | 2.5 | 0.040 |
| F | 5 | 5 | 0.046 |
| G | 2.5 | 7.5 | 0.051 |
| H | 7.5 | 7.5 | 0.066 |

**Table 2**

| Sample | Nominal C / B ratio | $a$ (Å) | $c$ (Å) |
|---|---|---|---|
| A | 0 | 3.0854(8) | 3.5239(10) |
| B | 0.019 | 3.0815(2) | 3.5275(3) |
| C | 0.036 | 3.0797(5) | 3.5275(6) |
| D | 0.043 | 3.0795(4) | 3.5283(5) |
| E | 0.040 | 3.0751(5) | 3.5273(9) |
| F | 0.046 | 3.0736(7) | 3.5261(8) |
| G | 0.051 | 3.0737(6) | 3.5265(7) |
| H | 0.066 | 3.0711(7) | 3.5263(9) |

**Figure Captions**

**Fig. 1.** Lattice parameters vs. nominal C/B ratio for the wires with $B_4C$ additives (stars) and with $B_4C$+SiC additives (circles).

**Fig. 2.** $J_c$ vs. $B$ for Fe/MgB$_2$ wires containing 0, 5, 10 and 12.5 wt.% $B_4C$ additions, reacted 1 hour at 800°C.

**Fig. 3.** $J_c$ vs. $B$ for Fe/MgB$_2$ wires containing the additives $B_4C$ and SiC in the proportions 7.5/2.5, 5/5, 2.5/7.5 and 7.5/7.5, respectively (all numbers are wt.% with respect to MgB$_2$).

**Fig. 4.** $J_c$ vs. $B$ for wire E with 7.5 wt.% $B_4C$ + 2.5 wt.% SiC additions at 4.2 and 20 K. The data of Soltanian et al. [4] have been added for comparison.

**Fig. 5.** Resistive transitions vs. B (inset) and the corresponding values of $B_{c2}$ and $B_{irr}$ up to 10T.

**Fig. 6.** $T_c$ of MgB$_2$ wires (measured by means of SQUID) with ternary ($B_4C$) and quaternary ($B_4C$ + SiC) additions as a function of the lattice parameter $a$.

**Figure 1**

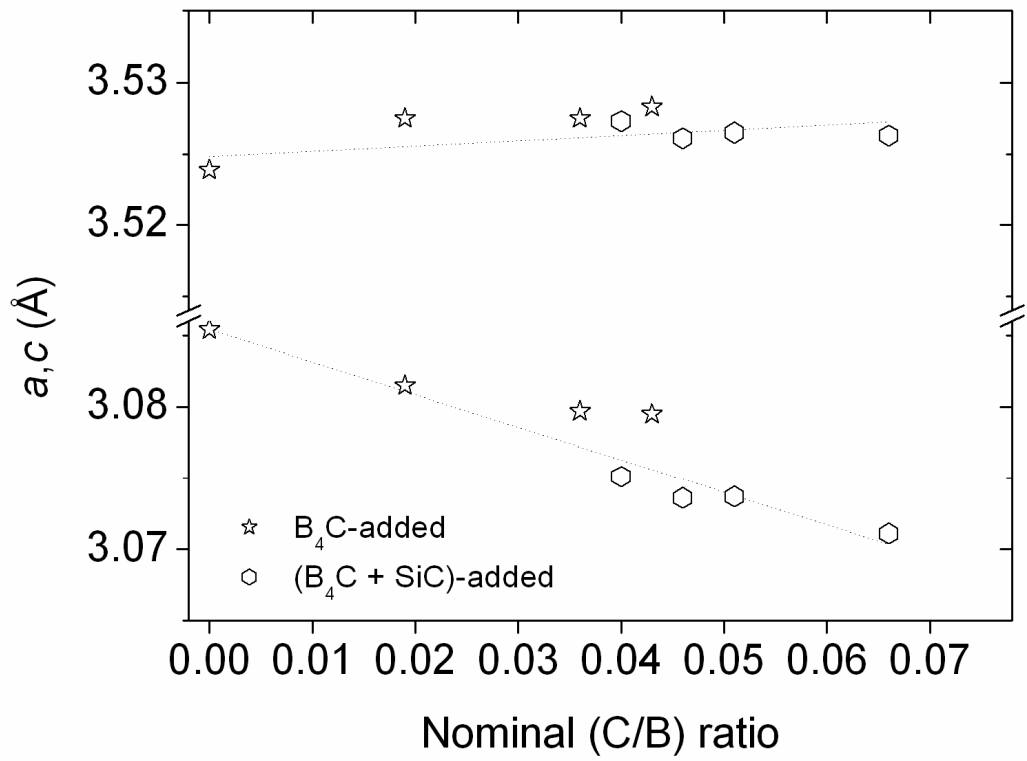

**Figure 2**

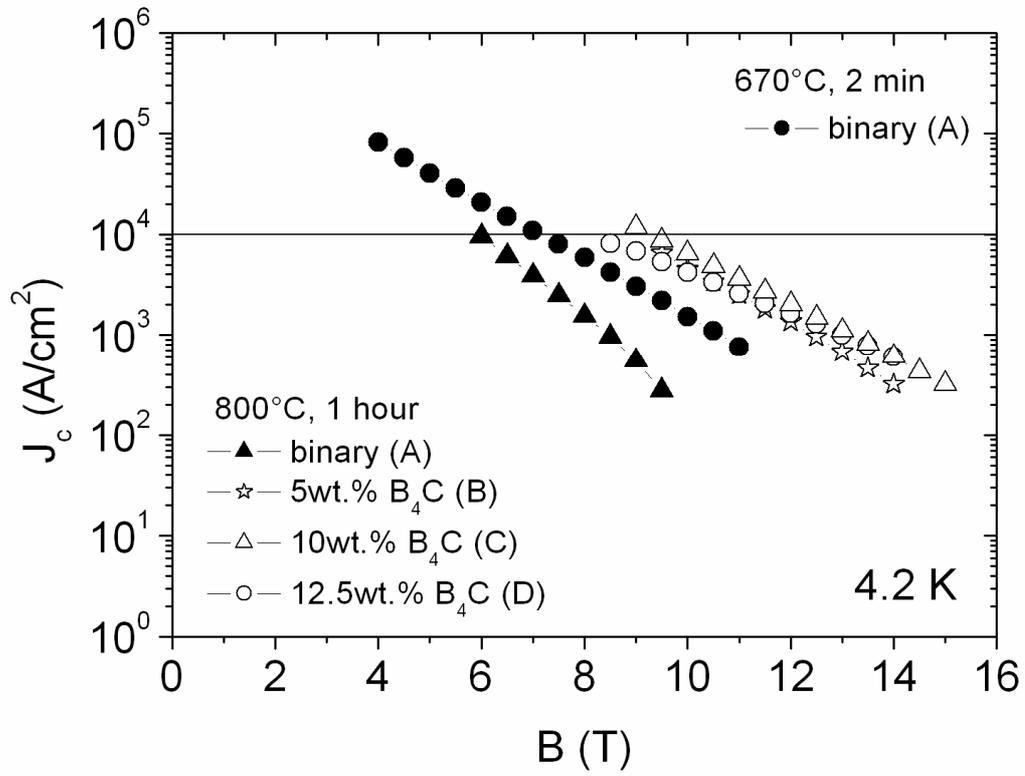

**Figure 3**

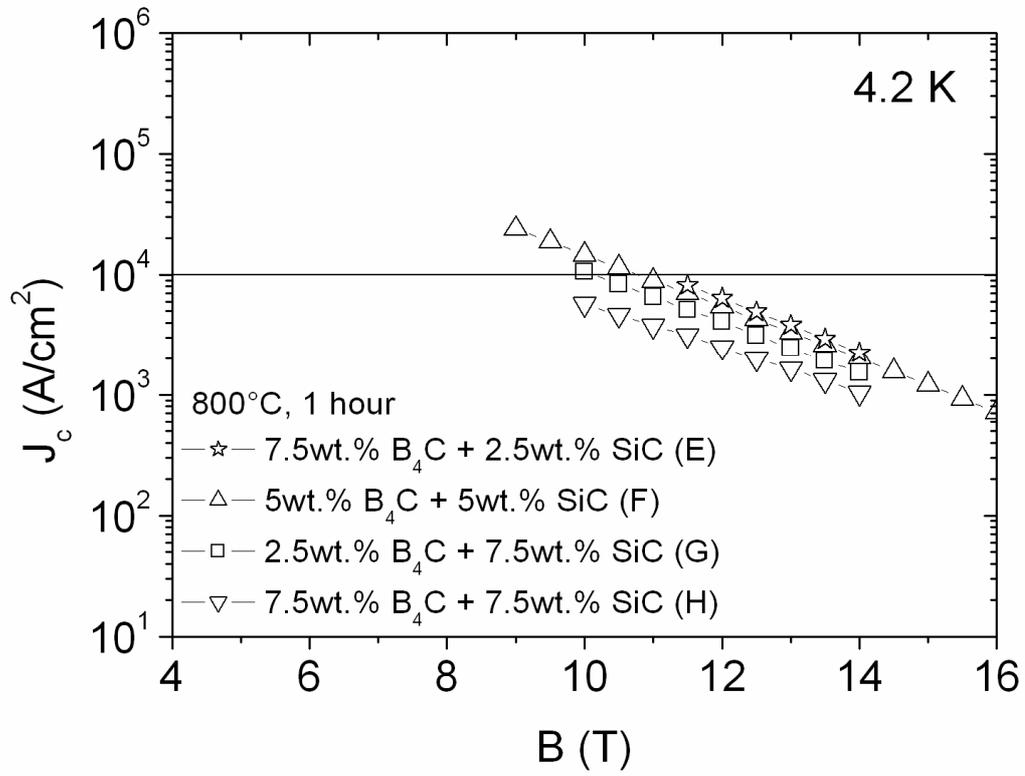

**Figure 4**

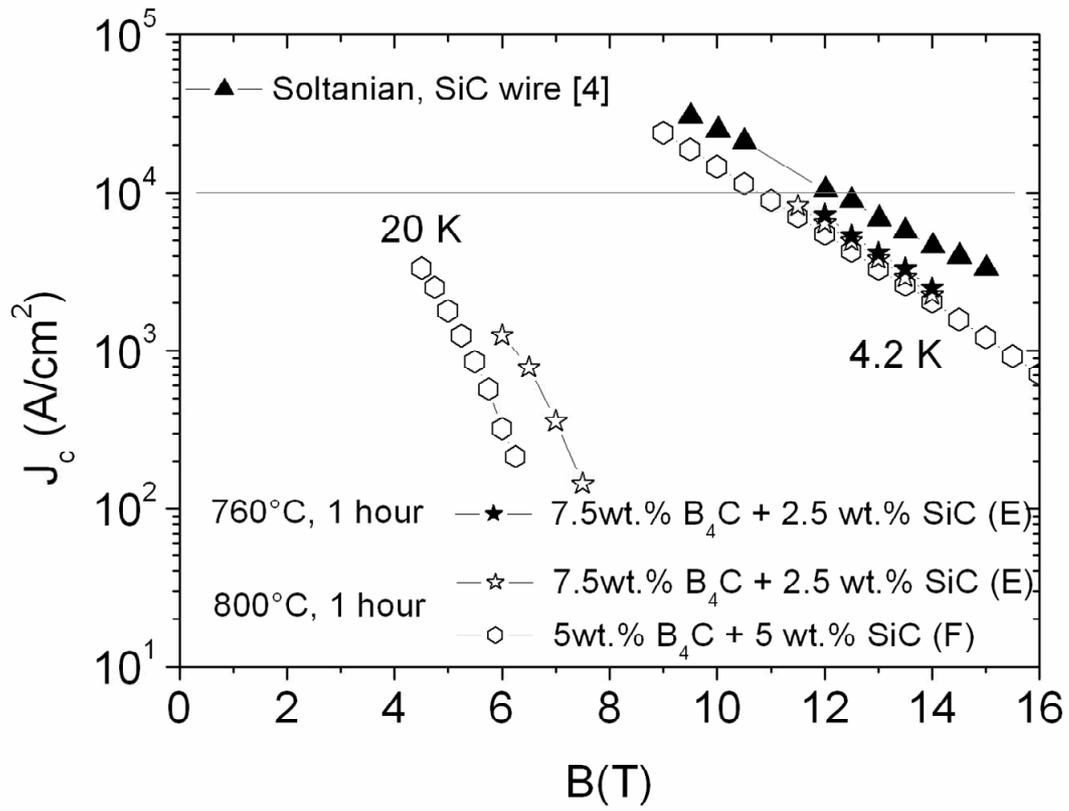

**Figure 5**

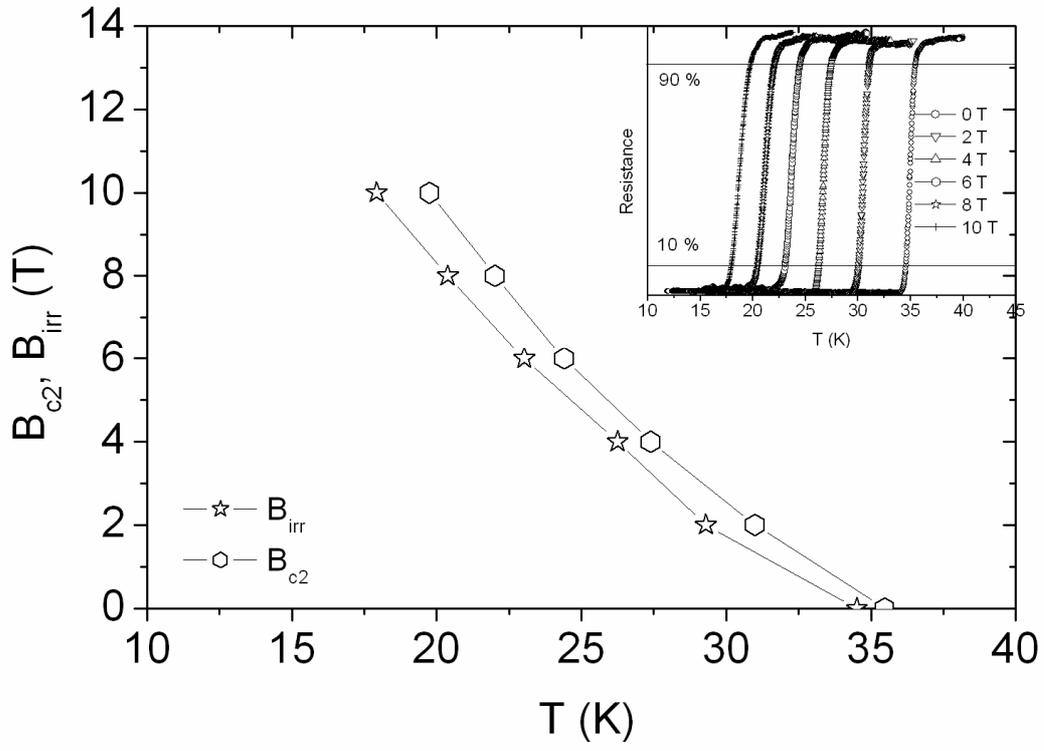

**Figure 6**

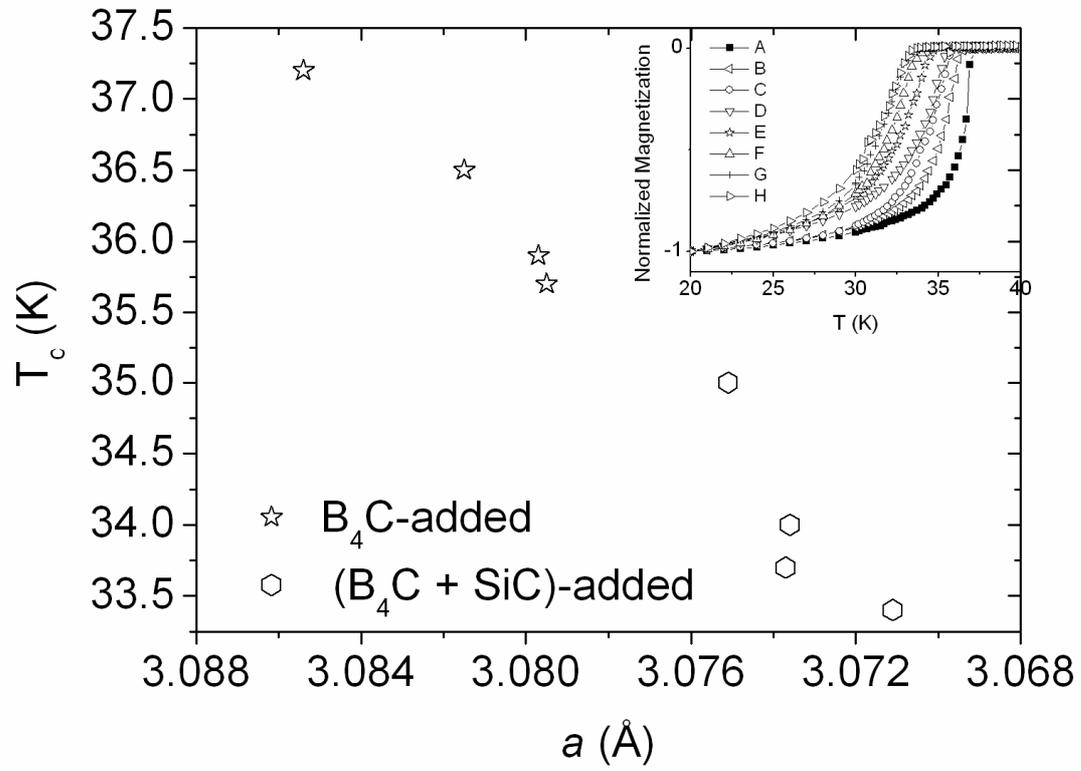